%% file: JMLFS_template.tex
\documentclass{JMLFS}

\def\be{\begin{equation}}
\def\ee{\end{equation}}
\def\bea{\begin{eqnarray}}
\def\eea{\end{eqnarray}}

\usepackage{placeins}
\usepackage{amsmath}
\usepackage{rotating}
\usepackage{subfigure}

\usepackage{graphicx}
\begin{document}

\title{\textit{SSEGEP}: Small SEGment Emphasized Performance evaluation metric for medical image segmentation}

\author{Ammu R\auno{1}, Neelam Sinha\auno{2}}
\address{$^1$International Institute of Information Technology,
Bangalore, India}
\address{$^2$International Institute of Information Technology,
Bangalore, India}

\begin{abstract}
Automatic image segmentation is a critical component of medical image analysis, and hence quantifying segmentation performance is crucial. Challenges in medical image segmentation are mainly due to spatial variations of regions to be segmented and imbalance in distribution of classes. Commonly used metrics treat all detected pixels, indiscriminately. However, pixels in smaller segments must be treated differently from pixels in larger segments, as detection of smaller ones aid in early treatment of associated disease and are also easier to miss. To address this, we propose a novel evaluation metric for segmentation performance, emphasizing smaller segments, by assigning higher weightage to smaller segment pixels. Weighted false positives are also considered in deriving the new metric named, "\textit{SSEGEP}"(Small SEGment Emphasized Performance evaluation metric), (range : 0(Bad) to 1(Good)). The experiments were performed on diverse anatomies(eye,liver,pancreas and breast) from publicly available datasets to show applicability of the proposed metric across different imaging techniques. Mean opinion score (\textit{MOS}) and statistical significance testing is used to quantify the relevance of proposed approach. Across 33 fundus images, where the largest exudate is 1.41\%, and the smallest is 0.0002\% of the image, the proposed metric is 30\% closer to \textit{MOS}, as compared to Dice Similarity Coefficient (\textit{DSC}). Statistical significance testing resulted in promising p-value of order $10^{-18}$ with \textit{SSEGEP} for hepatic tumor compared to \textit{DSC}. The proposed metric is found to perform better for the images having multiple segments for a single label. 
\end{abstract}

\maketitle


\section{Introduction}

Medical images need to satisfy a certain diagnostic quality criteria in order to obtain reliable information from them. The array of available quality metrics has been studied by Allister et al. \cite{Paper6}, as applied to Magnetic Resonance (MR) images. Subsequently the images could be subjected to various processing, of which image segmentation is very important. The segmentation is done in different ways depending on the image modality and characteristics of the region to be segmented. Some common methods include histogram analysis, region segmentation, and edge detection. Universal Network (U-net) based deep learning methods are also widely used in medical image segmentation. The subsequent steps that follow segmentation include feature extraction and classification based on the considered task. Hence the quality of segmentation severely affects the proceeding stages and thereby arises the need of robust performance evaluation metrics.

\noindent
{\bf General image segmentation and medical image segmentation:}
Evaluation of medical image segmentation should not be done similar to the general segmentation scenarios. State-of-the-art metrics include \textit{Accuracy}, \textit{Sensitivity}, \textit{Specificity}, Dice Similarity Coefficient (\textit{DSC}), Intersection Over Union (\textit{IOU}) and boundary distance scores which are applicable to general segmentation scenarios. Information about overlap area between ground truth and segmented image is used to evaluate the quality of general image segmentation method. But medical image segmentation evaluation should also consider the diagnosis aspect, like what is more important in disease diagnosis. In certain cases, contour of the segment is of interest which requires the segmentation algorithm to provide exact boundary delimitation. But in other cases, location of the segment may be of interest. There may also be cases where the size/volume of the segment need to be analyzed.  Some examples of medical segmentation scenarios include the detection and monitoring of tumor progress which is important in surgical planning\cite{p10,p11,p12}, change patterns in white matter that assists in early diagnosis and treatment of brain diseases\cite{p13}. Other anatomic features in medical image segmentation include gray matter, lesions, organs and tumors. There are also medical scenarios where there are smaller segments that convey more information on disease grading/diagnosis.  

Hence, the requirements on the segmentation algorithms for medical images depend on the goal of segmentation. Different evaluation metrics are sensitive to different properties of segmentations. Therefore, the evaluation metric should be chosen based on that. If only boundary of the region needs to be extracted, boundary-based distance scores can be used for evaluation. If it is required to know the total area/volume of a single tumor present in a tissue, \textit{DSC} or \textit{IOU} can be used for evaluation.

\begin{figure}
      \centering
      \includegraphics[scale=0.09]{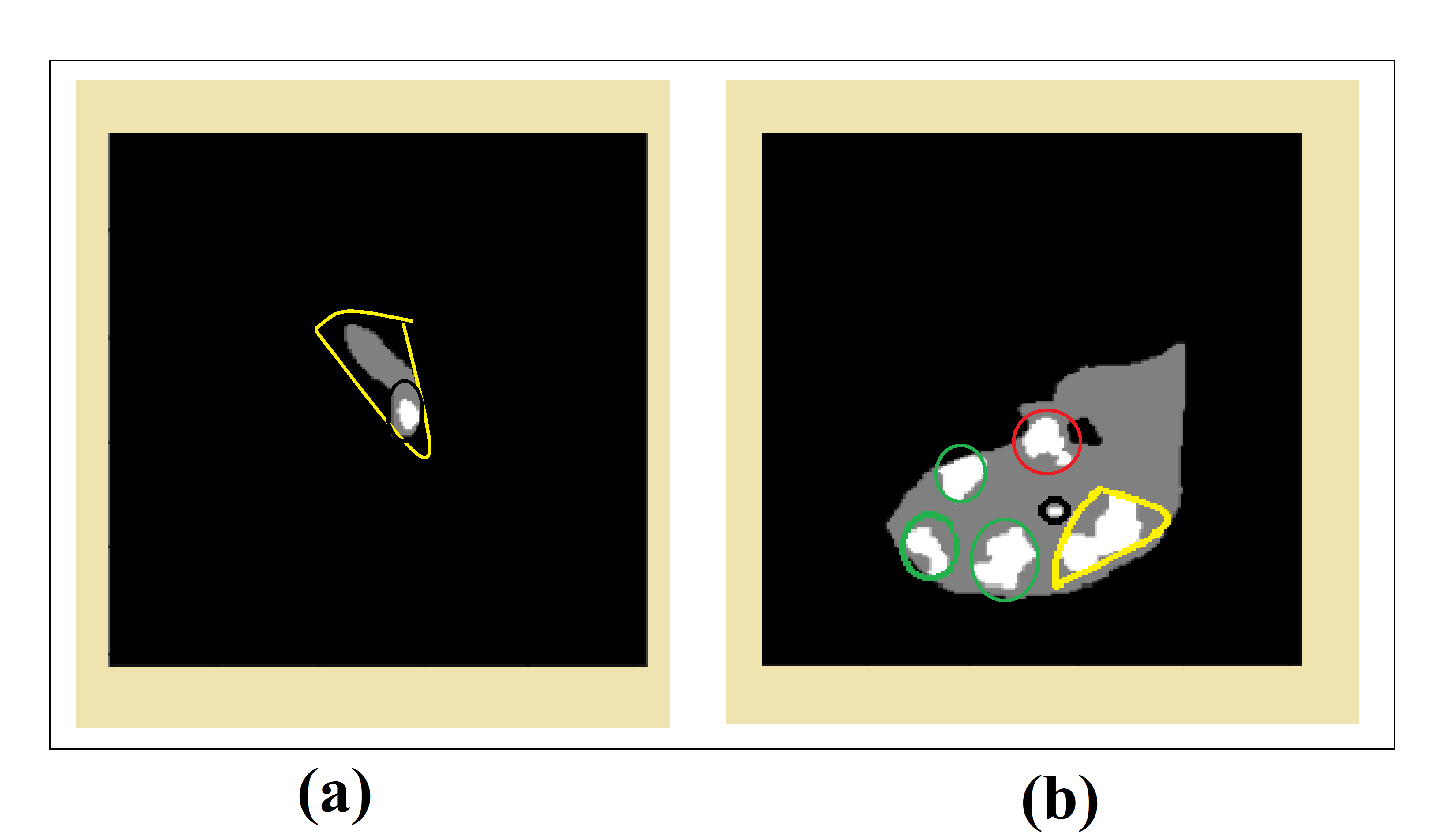}
      \caption{(a) Multi-label segmentation: Pancreas(circled yellow) and tumor(circled black). (b) Multi-label segmentation with tumors of varying sizes. Tumors in (b) are shown enclosed in colored contours.}
      \label{lab:multilab}
\end{figure}

\noindent
{\bf Segmentation challenges in medical images:} The region to be segmented in medical images is not always distinct and localized. There may be regions with variation in location, size, shape and brightness. Preprocessing techniques usually take care of brightness and color. Other challenges include:
\begin{enumerate}
\item Imbalance in data distribution: When the size of the segment is significantly smaller than the background, then the metrics based on calculation of True Negatives {\sl (TN)} are not suitable. The number of \textit{TN} pixels is always larger in medical images as shown in Figure~\ref{lab:moretn}.
{\sl Accuracy},{\sl sensitivity}, {\sl specificity} and Positive Predictive Value {\sl (PPV)} are not suitable for pixel level segmentation evaluation since the True Negative {\sl (TN)} pixels which constitutes the major portion of the image is treated similar to that of other pixels. Pixel {\sl accuracy} and {\sl specificity} is usually high for medical images since {\sl TN} is more. Quality of segmentation can't be measured only with {\sl sensitivity} as it does not consider the False Positives {\sl FPs}. As {\sl PPV} considers only the positive pixels, it can't be used to asses the overall segmentation quality. So such classical metrics are not efficient for segmentation evaluation. {\sl DSC} (Also known as F1-score) and {\sl IOU} can be used in this case to overcome the above stated drawback as it considers the overlap between ground truth and prediction.
    \begin{figure*}
      \centering
    \includegraphics[width=6.5in]{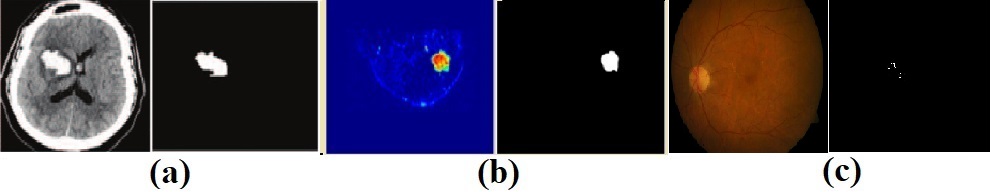}
      \caption{(a)Brain tumor CT slice(images taken from \cite{brainct}, (b) Dynamic Contrast Enhanced (DCE-MRI) slice with color map and tumor~\cite{mri_breastim}, (c) Fundus images(image taken from DIARETDB1 dataset). In each of the image pairs, the image on the left is the original, while the image on the right is the segmentation ground truth hand marked by experts.
      The fraction of pixels in the ROI, in each of the cases, is very small. Number of \textit{TN} pixels is large in all these  cases compared to the \textit{TP} pixels.}
      \label{lab:moretn}
\end{figure*}
\FloatBarrier
\item Variation in size: Another challenge faced in medical segmentation is due to the size variation of the regions to be segmented. The region may appear clustered and not localized at a point. Here, the challenge of segmentation algorithm is to extract the tiniest segment. Smaller segments usually indicate the presence of associated disease at an earlier stage and hence it is required to emphasise these small sized segments more compared to larger ones. State-of-the-art metrics like {\sl sensitivity}, \textit{specificity}, {\sl DSC} and {\sl IOU} does not discriminate the segments based on the area. So it is imperative to design a metric that highlights the detection of smaller segments. 
\begin{figure}
  \centering
  \includegraphics[scale=0.07]{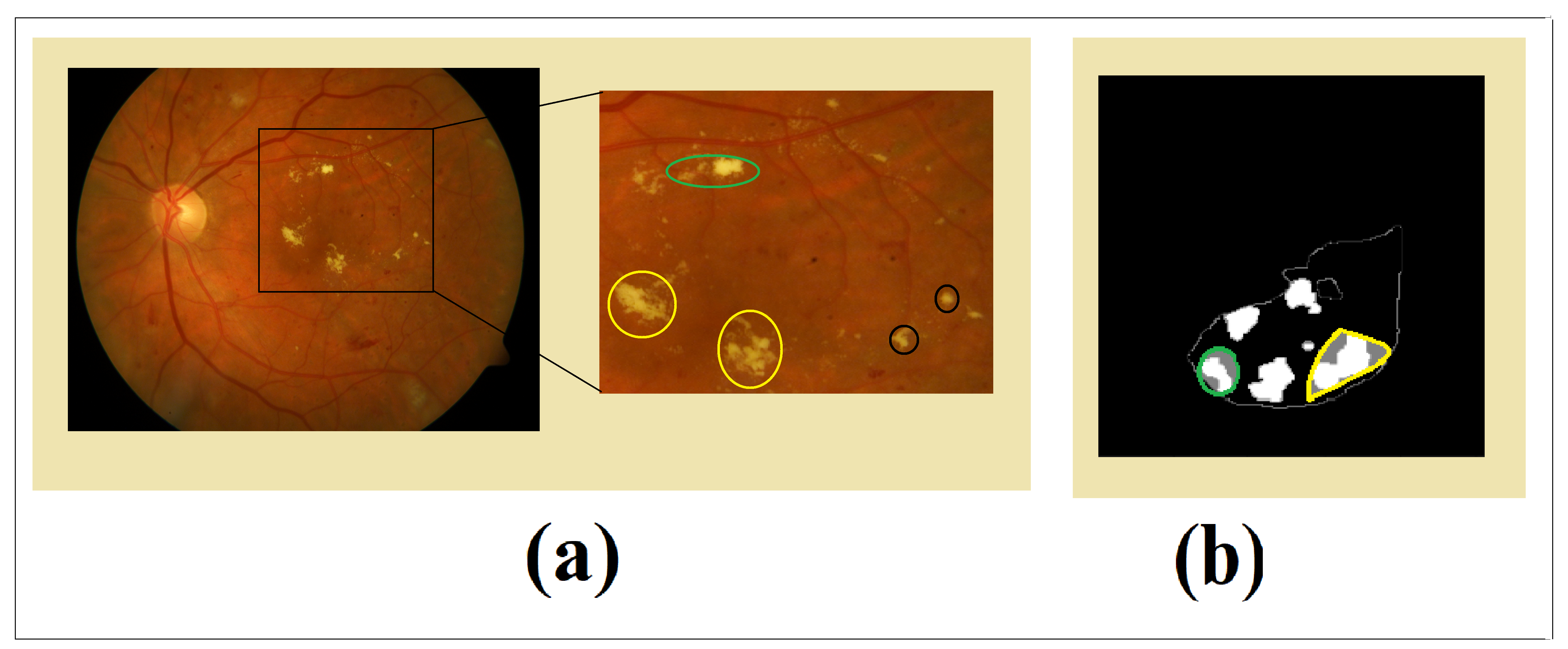}
  \caption{(a)Fundus image with exudates of various sizes circled in different colours. (b) Tumors in liver tissue of varying sizes.}
  \label{lab:size}
\end{figure}
Such a scenario can be seen in fundus images with exudates of various sizes as shown in Figure~\ref{lab:size}(a). A small protruding exudate may aid in early detection of diabetic retinopathy. Hepatic tumor is another example with multi-tumor case as seen in Figure~\ref{lab:size}(b). Another scenario is masses in mammograms which help to identify breast cancer.
\item Multi-label segmentation with class imbalance: Multi-label segmentation arises when there are segments of different labels in a single image and class imbalance occurs when there is a relative size difference in segments of each class(label). For example, the image in Figure~\ref{lab:multilab}(a) consists of pancreas tissue and a tumor(which occupies a smaller space compared to pancreas) within that. The loss function of deep learning network(for segmentation) favours the large sized pancreas more than the tumor. But the loss function should be efficient to extract both the segments. Commonly used loss function for medical segmentation is Dice loss which is derived based on the evaluation metric {\sl DSC}. But here, generalized Dice loss\cite{multidice} which works for multi label case should be used as it is not a simple binary segmentation. Another medical scenario is multi-label segmentation with more than one segment of various sizes in each category of label. Liver and tumor segmentation(Figure~\ref{lab:multilab}(b)) illustrates this situation with liver tissue and multiple hepatic tumors as the targets. In such a scenario, multi-label {\sl DSC} is not sufficient to quantify the segmentation performance, as there may be segments of different sizes in each category of label and it is required to penalize the loss incurred by each pixel based on the area of the segment it belongs to.
\end{enumerate}
\noindent
\textbf{Importance of small-sized segments in medical images}
Exudates are patches of fatty deposits on the eye formed due to leaky blood vessels. They appear as yellow-white structures in color fundus images with variable sizes, shapes and contrast. As exudates are among the common early clinical signs of Diabetic Retinopathy (DR), \cite{DR1} their detection would be critical in diagnosing the health status of the eye. The challenges faced in exudate segmentation include drastic variations in size(small to large as optic disc) and shape, and the severe imbalanced distribution of classes(more number of non-exudate pixels). The size of diabetic hard exudates changes with the progression of disease \cite{evidencefundus}, hence the identification of tiny exudates is very important as it helps in the early treatment. So, the performance evaluation should be in such a way that it emphasizes the smaller exudates more. As the commonly used metrics treat all the {\sl TP} pixels in similar way, they are not capable to quantify the small exudate segmentation. \cite{exudates1} have used {\sl sensitivity} and {\sl PPV} for exudate segmentation evaluation. \cite{exudates2} have used \textit{sensitivity}, {\sl specificity}, {\sl accuracy} and area under curve of Receiver Operating Characteristic (ROC) as metrics to evaluate the detection of hard exudates using mean shift and normalized cut method.

Another medical segmentation scenario is in cancer detection. To detect cancer with Computer Aided Detection (CAD) techniques, tumor segmentation is inevitable. Tumors are visible in the Computed Tomography (CT) scans taken by the radiologist and it appears in various sizes. Pancreatic tumor is smaller in size compared to the large pancreatic tissue and a common algorithm for multi-label segmentation(pancreatic tissue and tumor) is a challenge faced in segmentation. The loss function used in training deep networks should overcome this imbalance due to size difference. 
Most recent works use cross entropy and generalized {\sl DSC}\cite{multidice} in this scenario. Numerous lesions of varying sizes may appear in a liver tissue called as hepatic lesions which can be malignant. Studies \cite{smallhepaticlesions} have revealed the prevalence of small hepatic lesions found at CT in patients with cancer. These small lesions represent metastases in 11.6\% patients as identified by the research study. The size of the lesions also changes during course of cancer and patients with small hepatic metastases also had associated extra hepatic metastases. 

Calcifications are the earliest signs of breast cancer. Information about the size, density, and distribution of breast microcalcifications can give an idea about the benign or malignant nature of the cancer. It is found that decreasing calcifications is not a benign finding on mammogram\cite{disapp_cal}. Spontaneous decreases of calcifications are not necessarily indicative of better clinical outcomes. It has been demonstrated that a decrease in or complete resolution of breast calcifications is most concerning when it is associated with an extra breast mass, architectural distortion, or increased density\cite{calref1,calref2}. Small clusters of punctate or granular calcifications may represent high-grade {\sl DCIS}(Ductal carcinoma in situ), where an aggressive clinical approach is recommended\cite{calref3}. If the calcifications are larger in size and well defined in shape, then it is less suspicious. But if they appear smaller in size or appear in varying shapes, then they are cancerous\cite{calref4}. Hence, it is important to detect the smaller calcifications in mammograms.

The medical scenarios explained above states that the detection of smallest lesion is very critical in further medical analysis. The smallest exudate in Figure~\ref{lab:size}(a) occupies only 0.0002\% of the image, the smallest tumor in Figure~\ref{lab:multilab}(a) and (b) occupies 0.0015\% and 0.00004\% of the respective images. A small protruding lesion may indicate the start of a disease. In most of the cases, one of the factors which is used to grade the disease severity is size/volume of the lesion or region of interest. As explained before, size of the hard exudates increase during the course of diabetic retinopathy\cite{evidencefundus}, lesions become bigger with stages of cancer like hepatic cancer
\cite{smallhepaticlesions} and breast cancer. Hence, the detection of a tiny lesion helps in the early treatment of associated disease which can thereby avoid unnecessary biopsies. Also the detection of a small liver tumor may point the presence of a tumor in another region\cite{smallhepaticlesions}. These studies enunciate that a performance evaluation metric that emphasizes these tiny lesions is imperative. 

Olaf Ronneberger \cite{unet1} used UNet convolutional network for biomedical image segmentation. Keyvan Kasiri et al. \cite{keyvan} used {\sl DSC} to compare three brain MRI segmentation methods in software tools. A recent work reported by Ona et al. \cite{paper1} proposed an automated method to segment ischemic lesions in Diffusion-weighted MRI images using an ensemble of 3D convolutional neural networks (CNN). {\sl Dice} score was utilized to evaluate the segmentation performance. The work reported by Piantadosi et al. \cite{paper2} achieves segmentation of breast parenchyma on volume MR data, utilizing an ensemble of CNNs. {\sl Dice} score was utilized to quantify the segmentation {\sl accuracy}. In the work reported by Ting et al. \cite{paper4}, the authors propose to improve segmentation accuracy by incorporating representation using Sparse Dictionary-based Learning, in the context of lesion segmentation in brain MR images. Nur et al. \cite{paper8} used saliency method based on region to detect exudates in retinal images of Diabetic Retinopathy. They have calculated {\sl accuracy}, {\sl sensitivity} and \textit{specificity} for the evaluation of the proposed method. Abblin et al. \cite{paper9} calculated {\sl accuracy} to evaluate the segmentation and recognition of iris from eye image. The authors used Fuzzy interference Logic based edge detection technique for segmentation. The work reported by Priya et al. \cite{paper10} used support vector neural network classifier for object and image level classification of Tuberculosis images and {\sl accuracy} was calculated for evaluation. Fang Lu et al. \cite{paper11} used volumetric overlap error {\sl (VOE)}, relative volume diﬀerence (\textit{RVD}), average symmetric surface distance (\textit{ASD}), root mean square symmetric surface distance (\textit{RMSD}) and maximum symmetric surface distance (\textit{MSD}) to evaluate the deep learning algorithm with graph cut refinement to automatically segment liver in CT scans. Peter rot et al. \cite{paper12} reported work on deep sclera segmentation and recognition in which \textit{precision}, \textit{recall} and \textit{F1-score} were used for segmentation evaluation and False Acceptance (\textit{FAR}) and False Rejection error Rates (\textit{FRR}) and Area Under the ROC curve (\textit{AUC}) scores for the evaluation of recognition methods.
An entirely different objective of comparing patch-based versus volumetric analysis of MR images was studied by Bontempi et al. \cite{Paper3}. Here, statistical significance testing was carried out to illustrate the superiority in segmentation performance while using entire volumes rather than patch-based analysis. The probability maps produced by the 2D and 3D-patch based methods are characterised by the presence of voxels associated with significant probability of belonging to the structure of interest (p > 0.2) despite their distance from the latter. In the work reported by \cite{paper7}, the authors propose a method to segment structures in brain MR in Unsupervised setting, using \textit{Dice} score and the Area under the \textit{Precision-Recall} curve as the quantitative measures. \cite{dice1,dice2} also used \textit{DSC} as an evaluation metric in medical image segmentation. \cite{multidice} have used Generalized \textit{Dice} overlap for segmentation assessment for unbalanced tasks. A reported work by Nabila et al. \cite{paper5}, discusses the issue of class imbalance in medical images. The authors propose generalized focal loss function in order to evaluate the segmentation performance on small-sized segments, using CNNs. However, in this approach, authors tweak the cost-function during the Training phase, in order to obtain better segmentation. \cite{familymetrics} have introduced a family of metrics to quantify the difference or similarity of segments by considering their average overlap in fixed-size neighborhoods of points on the boundaries of those segments.  

In order to quantify the segmentation performance of images with varying sizes of segments(exudates or tumors), we require a metric which emphasizes the smaller-sized segments. As seen before, \textit{DSC} assigns the same weightage to all the \textit{TP} pixels irrespective of whether it is part of large/small region, which seems inappropriate. The small segments should be penalized more while calculating the segmentation loss since the loss of few number of pixels in a small segment is severe than that in a large segment. The existing metrics as far as our knowledge, does not discriminate the detection of larger and smaller segment pixels. \textit{DSC} and \textit{IOU} metrics reflect the human understanding when the data set contains images with only large or only small segments with less false positives. For a clinical evaluation where screening has to be done, image level evaluation using \textit{sensitivity} and \textit{specificity} is enough. But to compare different segmentation methods or to optimize a single method, metrics that consider overlap between ground truth and segmented image is required. As the medical data sets usually consists of segments of varying sizes,
evaluation with \textit{DSC} and \textit{IOU} is not sufficient as they treat all segments in the same way. The proposed metric which incorporates the area of segments and which penalizes the loss of smaller segments more would be more effective in comparing the segmentation approaches.
This paper is organized as follows: section II discusses the commonly used
evaluation metrics, section III presents the proposed approach in detail, section IV demonstrates the experimental results and analysis with the data sets used, followed by conclusion in section VI.

\section{Existing Evaluation Metrics}
Different evaluation metrics are used in practice to compare the effectiveness of various segmentation algorithms. The evaluation metric to be used depends on the problem and anatomy of the segmented region. To compare various medical segmentation methods, we need evaluation at pixel level for the entire image. The metric definitions of most popular evaluation methods are given in Table~\ref{tab:eq}.
\begin{table}[ht!]
\tbl{Most popular evaluation metrics and their definitions are given. '$|$ $|$' represents the cardinality. G and S constitutes the set of non-zero pixels in ground truth and prediction respectively. $|I|$=TP+FP+FN+TN, '$l$' is the label length in multi-label segmentation.\label{tab:eq}}{%
\begin{tabular}{cc}
\toprule
Metric  & Formula \\
\colrule
Accuracy & \(\displaystyle
Acc=\frac{|I|-|G\cup S|+|G\cap S|}{|I|}\)\\
Sensitivity & \(\displaystyle
    Sens=\frac{|G\cap S|}{|G|}\)\\
Specificity & \(\displaystyle
    Spec=\frac{|I|-|S\cup G|}{|I|-|G|}\)\\
PPV & \(\displaystyle
    PPV=\frac{|G\cap S|}{|S|}\)\\
IOU & \(\displaystyle
    IOU=\frac{|G\cap S|}{|G\cup S|}\)\\
Hausdorff score & \(\displaystyle
    HD=\)\\
(Boundary based score) & \(\displaystyle\max\{{\max_{x\epsilon\partial G}{d(x,\partial S)}},{\max_{y\epsilon\partial S}{d(y,\partial G)}}\}\)\\ 
DSC & \(\displaystyle
    DSC=\frac{2|G\cap S|}{|S|+|G|}\) \\
Generalized Dice score & \(\displaystyle
    GD=2\frac{\sum_{i=1}^{l}\frac{1}{|Gi|}|Gi\cap Si|}{\sum_{i=1}^{l}\frac{1}{|Gi|}(|Si|+|Gi|)}
    \)\\
\botrule
\end{tabular}}
\end{table}

\noindent
An imbalance in data distribution is noticed in medical image segmentation where \textit{TN} pixels is higher compared to other pixels, which makes classical metrics like \textit{accuracy} and \textit{specificity} inappropriate for performance evaluation. 
\begin{figure*}
      \centering
      \includegraphics[scale=0.5]{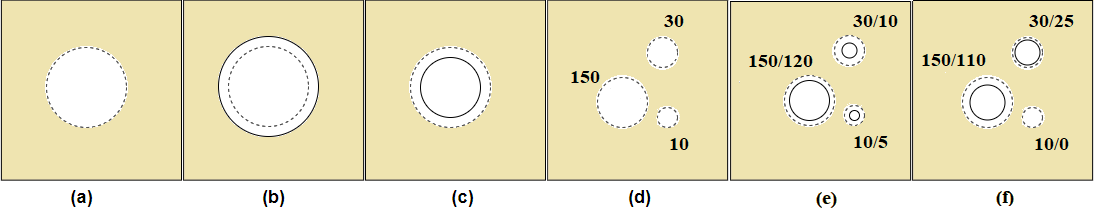}
      \caption{Synthetic images to illustrate the limitations of evaluation metrics:(Numbers shown in the figure indicate the number of pixels that make up the area of the nearest segment. Regarding fractions, the numerator denotes number of pixels in groundtruth segment and denominator denotes number of pixels in predicted segment) (1)Boundary distance based score and (2)Dice Similarity Coefficient. Ground truth is shown using dashed line and predictions in solid line. (a) Ground truth which is a single chunk. (b) and (c) depicts two different segmentation scenarios of the ground truth in (a). Prediction is a sub-set of ground truth in (b). Prediction is a super-set of ground truth in (c). Boundary distance based scores for (b) and (c) are same as the difference between the radius of the rings are equal. (d) Ground truth which is represented as diverse-sized interconnected segments. (e) and (f) are two segmentation scenarios of ground truth in (d). Same number of pixels are missed in segmentation for (e) and (f). (e):TP=120+10+5=135,FP=0,FN=55 (f) TP=110+25+0=135,FP=0,FN=55 leading to same \textit{DSC}.}
      \label{synth_im}
\end{figure*}
\noindent
Overlap metrics like \textit{DSC} and \textit{IOU} takes into account the area of overlap between the ground truth and segmented image. Both these metrics measure the same aspects and hence it is not required to use them together as evaluation metrics. Boundary distance based scores calculates the distance between the boundaries of segments in ground truth and prediction. They are used when boundary delineation of the segmentation is important. It does not take into account the size information of regions. Hausdorff Distance is not recommended to directly use in medical image segmentation as it is sensitive to outliers\cite{ref_hau} and time-consuming. Another interesting metric is \textit{MCC} (Matthews Correlation Coefficient)\cite{mcc_ref} which is symmetric in nature where no class is important than the other. It takes into account all four values in confusion matrix. It can be used when it is required to detect both the classes(when TN is also important) and/or classes are imbalanced.

If images in the data set consists of segments of various sizes, the segmentation algorithm should be able to detect most of the pixels in the smaller segment, even if it misses few larger segment pixels. So to assess the quality of prediction, the evaluation metric should be able to penalize more for smaller segment pixels than for larger segment pixels.  This aspect is not considered in simple overlap and boundary distance based metrics.
For the evaluation of multi-label segmentation performance as in pancreatic tumor, generalized \textit{Dice} score\cite{multidice} which gives a weightage of inverse of total area for each label is used. But if there are many segments in a single class(many hepatic lesions), \textit{DSC} is not sufficient for performance evaluation as it can't discriminate between the segments in a single class. 
Few synthetic images which represent different binary segmentation scenarios are constructed  as  shown  in  Figure~\ref{synth_im}. The distance between the boundaries of ground truth and prediction for the images in (b) and (c) are same even though two different cases are depicted, which shows the failure of boundary based metric. Since same number of pixels are lost in (e) and (f), \textit{DSC} will be the same for both, even though a full small segment is lost as seen in (f). This drawback is overcome by the proposed metric as it weighs the \textit{TP} pixels based on the area of the segment it is derived from. Higher weightage is given to the smaller segment pixels and lower weightage to the larger segment pixels. 

\section{Methodology}

\subsection{Proposed Metric}
We have proposed an evaluation metric to quantify the performance of various segmentation scenarios, especially the cases where small segments in the prediction need to be emphasized more. We have illustrated the analysis in medical scan images which contains segments of various sizes. The proposed metric, \textit{SSEGEP} penalizes the loss of smaller segments more in comparison to the larger ones. 
Each pixel in the ground truth image is assigned a weightage which is calculated based on the area of the segment it belongs to. Weightage of a segment is the inverse of its area, which results in high-weighted small sized segments and low-weighted bigger segments. So while calculating \textit{SSEGEP}, instead of simple count of true positive pixels, weighted sum of TPs (True Positive) is considered. \textit{FP} pixels are also weighted which is given as the inverse of \textit{TP} count. If the task given is to quantify the combined segmentation of organ and tumor, two different weights are given to the \textit{FP} pixels: one for tumor case, which is given as the inverse of \textit{TP} count of tumor mask and other for organ case, which is given as the inverse of \textit{TP} count of organ mask.


\subsection{Implementation and Explanation}
\begin{sidewaysfigure}
  \vspace{-2.5em}
    \centering
    \includegraphics[width=20.5cm]{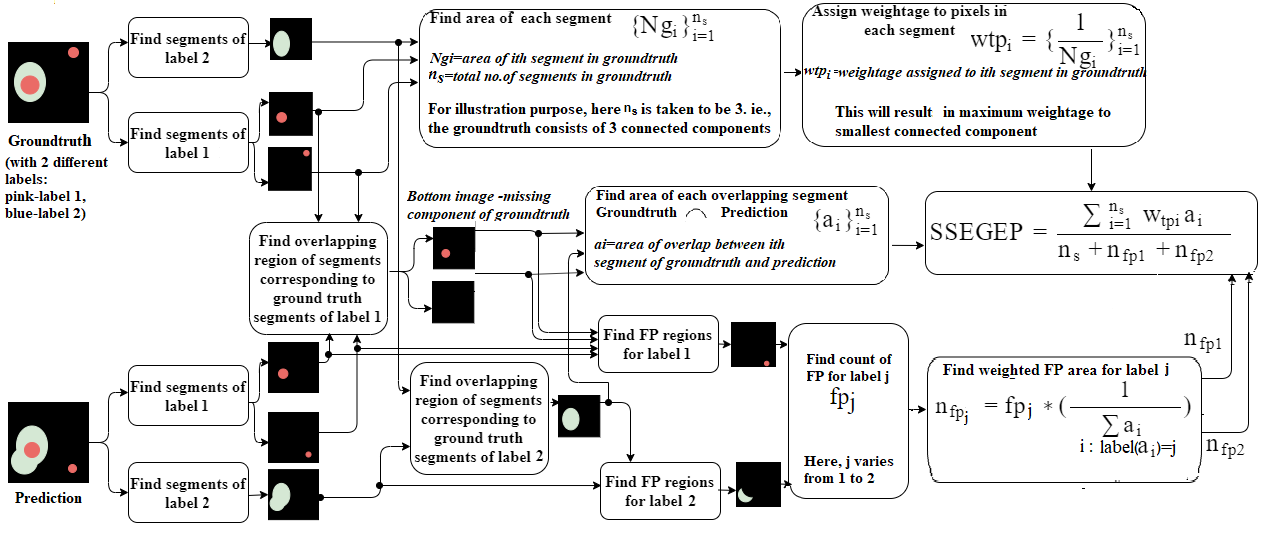}
    \caption{General outline of proposed evaluation approach: Weightage is assigned to \textit{TP} pixels and \textit{FP} pixels after finding the segments from the ground truth image to calculate the proposed metric. \textit{FN} pixels are implicitly taken care while assigning weightage to \textit{TP} pixels since more number of \textit{FN} pixels lead to small value of $a_i$ and lesser value of numerator in SSEGEP. Here, $n_s$ is the total number of segments(including label 1 and 2).}
    \label{fig:blkdgm}
      \vspace{-2.5em}
\end{sidewaysfigure}

The general block diagram for 2D images is shown in Figure~\ref{fig:blkdgm}. The ground truth and segmented images are given as inputs. Segments of label 1 and 2 are given in different colors. The segmentation evaluation methodology is comprised of assigning weightage to segments in ground truth image, finding true positive pixels in each segment, finding and assigning weightage to the false positive segments and final calculation of proposed metric. Initially, from the ground truth image, distinct segments belonging to each label is found. Here, in the block diagram, there are two different labels and every distinct segment of these two labels are found. Similarly, the distinct segments of each label are found from the segmented image. This is done by finding the contours of the input image. Then, a weightage is allotted to these ground truth segments which is given as the inverse of area of that segment as shown in Figure~\ref{fig:blkdgm}. Then, the count of pixels in \textit{TP} segments(overlapping region of segments between ground truth and prediction) is calculated which is given by $a_i$ as shown in the block diagram. XOR operation is performed separately for each label between the groundtruth and segmented image to find the false positive images for each label(Here, two \textit{FP} images are obtained as there are two labels), and the separate count of these pixels is recorded. Weightage of inverse \textit{TP} count is given to those \textit{FP} pixels as shown in the block diagram. So, when the count of \textit{FP} is very much higher than that of \textit{TP}, SSEGEP value becomes small indicating poor segmentation, and when the count of \textit{FP} is small compared to \textit{TP}, SSEGEP value becomes high indicating good segmentation(provided FN is less). Final \textit{SSEGEP} value is calculated from these set of metrics which quantifies the quality of the prediction. The proposed metric is given by equation~(\ref{eq:segp}).
\begin{equation}
\label{eq:segp}
    \textit{SSEGEP}=\frac{\sum_{i=1}^{n_s} {w_{tpi}} {a_i}}{n_s+\sum_{j}{n_{fpj}}}
\end{equation}

Where $w_{tpi}$ and $a_i$ represents the weightage and \textit{TP} count of $i^{th}$ segment respectively and $n_s$ represents the total number of segments in the ground truth image. Here, $i$ is used to indicate segment and $j$ to indicate label.

\begin{equation}
    n_{fpj}=fpj*(\frac{1}{\sum_{label(a_i)=j}{a_i}})
\end{equation}
 

Where $fp_j$ represents the total \textit{FP} count for segments of label j.



\section{Experimental Details and Materials Used}
\input{expdis2.tex}

\section{Results and Discussion}
\input{expdis3.tex}

\section{Conclusion}
The segmentation of smaller lesions that appear in medical image scans are important as they assist in the early treatment of associated diseases. The evaluation metric used should be able to quantify the segmentation performance emphasizing these smaller lesions. The proposed metric assigns weightage to each segment in the image based on their area in such a way to emphasize the smaller segments more. The applicability of our metric across different imaging techniques is shown by testing in four public datasets of different anatomies and its significance is proved with mean opinion score and hypothesis testing. It is observed that the proposed metric is 30\% closer to \textit{MOS}, as compared to
\textit{DSC} for fundus images and statistical significance testing resulted in promising p-value of the order $10^{-18}$ with \textit{SSEGEP} for liver tumor compared to Dice Similarity Coefficient. The experimental results show that the proposed metric outperforms other metrics with respect to human interpretability and understanding for those images with multiple segments of various sizes.

\section*{Acknowledgements}
We would like to thank Fabian Isensee for providing the segmentation results of pancreas and liver images that are obtained using nnU-Net~\cite{DBLP:journals/corr/abs-1904-08128}. We also acknowledge Sanjeev Dubey for providing detailed analysis in GitHub about the segmentation algorithm~\cite{inbook1} used for fundus images.

\bibliographystyle{unsrt}
\bibliography{cas-refs}
\end{document}

%% file: expdis2.tex
The experiments are performed on synthetic images to illustrate the proof of concept and real scan images of three different modalities: Fundus, mammogram and CT to analyse and compare the proposed metric in different scenarios with other state-of-the-art metrics. Segmentation evaluation was done on:

(a) Images with multiple segments from same class: The images may contain segments of various sizes in which smaller segment is crucial in disease diagnosis. For eg, liver contain multiple tumors, fundus image consists of exudates and mammograms with calcifications of various sizes.

(b) Images with multiple segments from two different classes: Usually a single algorithm may not be efficient to extract segments of different labels in multi-label segmentation. In scenarios where a common algorithm is required to perform multi-label segmentation where the smaller segments are more important in disease diagnosis, it is imperative to use an evaluation metric which discriminates the segments in such a way weighting smaller segments more. As the proposed metric \textit{SSEGEP} assigns size-based weight to the segments in the input image, smallest segment gets the highest importance and the larger one gets the least weightage.

Further, the relevance of proposed metric, \textit{SSEGEP} is analysed with two different approaches: 

(1) \emph{Mean opinion score}: The proposed metric is compared with \textit{TPR} and \textit{Dice} score and its relevance is tested with mean opinion score calculated based on the scores assigned by 15 subjects on the 33 fundus segmentation results obtained. The scores were given in the range of (0,0.25,0.5,0.75,1) where 0 and 1 indicating poor and best segmentation respectively. Then the average
of the difference between mean opinion score and value of the evaluation metric is found which indicates the deviation of that metric from the mean opinion score.

(2) \emph{Hypothesis testing}: As we have a greater number of images in liver, pancreas and mammogram data sets, statistical significance of \textit{SSEGEP} was measured separately for each data set by hypothesis testing. To assess the statistical significance of the proposed metric for mammogram database, Welch's t-test was performed  since Levene’s test for homogeneity of variances indicated unequal variances between groups and sample sizes were unequal. Two tests were performed for a single data set: one for \textit{Dice} value and other for \textit{SSEGEP}. The inputs for the test are metric values for very good and very poor segmentation(extreme cases are considered for better understanding).The threshold to TP and FP to decide the samples in two groups were selected as:  If \textit{TPR} is greater than 80\% and \textit{FPR} and \textit{FNR} are lesser than 20\%, then the corresponding segmented image belongs to 'good segmentation' category. If \textit{TPR} is less than 40\% and \textit{FPR} and \textit{FNR} are more than 50\%, then the image belongs to 'bad segmentation' category. To conduct the test for \textit{SSEGEP}, the \textit{SSEGEP} values for 'good segmentation' category forms one input and the \textit{SSEGEP} values for 'poor segmentation' category forms the other input. Significance level ($\alpha$) is set to be 0.00001. If the resulting p-value is lesser than the set significance level, then the data is said to be statistically significant, which means that two populations have unequal means. The same test is repeated for \textit{Dice} to obtain the p-value.

All metric values \textit{Dice}, \textit{Accuracy}, \textit{IOU} and \textit{SSEGEP} – lie in the range 0 to 1 Python 3.6 is used for the implementation.
\subsection{Datasets Used}
We have used four different publicly available data sets to evaluate the significance of our metric. They are-
\begin{enumerate}
    \item DIARETDB1 - containing fundus images (Imaging modality = Fundus, anatomy = eye)
    \item CBIS-DDSM - containing mammogram images (Imaging modality = Mammogram, anatomy = breast)
    \item MSD challenge - containing liver images (Imaging modality = CT, anatomy = liver)
    \item MSD challenge - containing pancreas images (Imaging modality = CT, anatomy = pancreas)
\end{enumerate}
They are all publicly available databases, with 89 images of the eye, 131 training images of the liver, 282 training images of pancreas, 173 training images of the breast with calcification upon which several research groups have evaluated their algorithms. Overall, 379 images from real scans, were used to illustrate the performance of the proposed metric.

\begin{figure}
      \centering
      \subfigure{\includegraphics[scale=0.075]{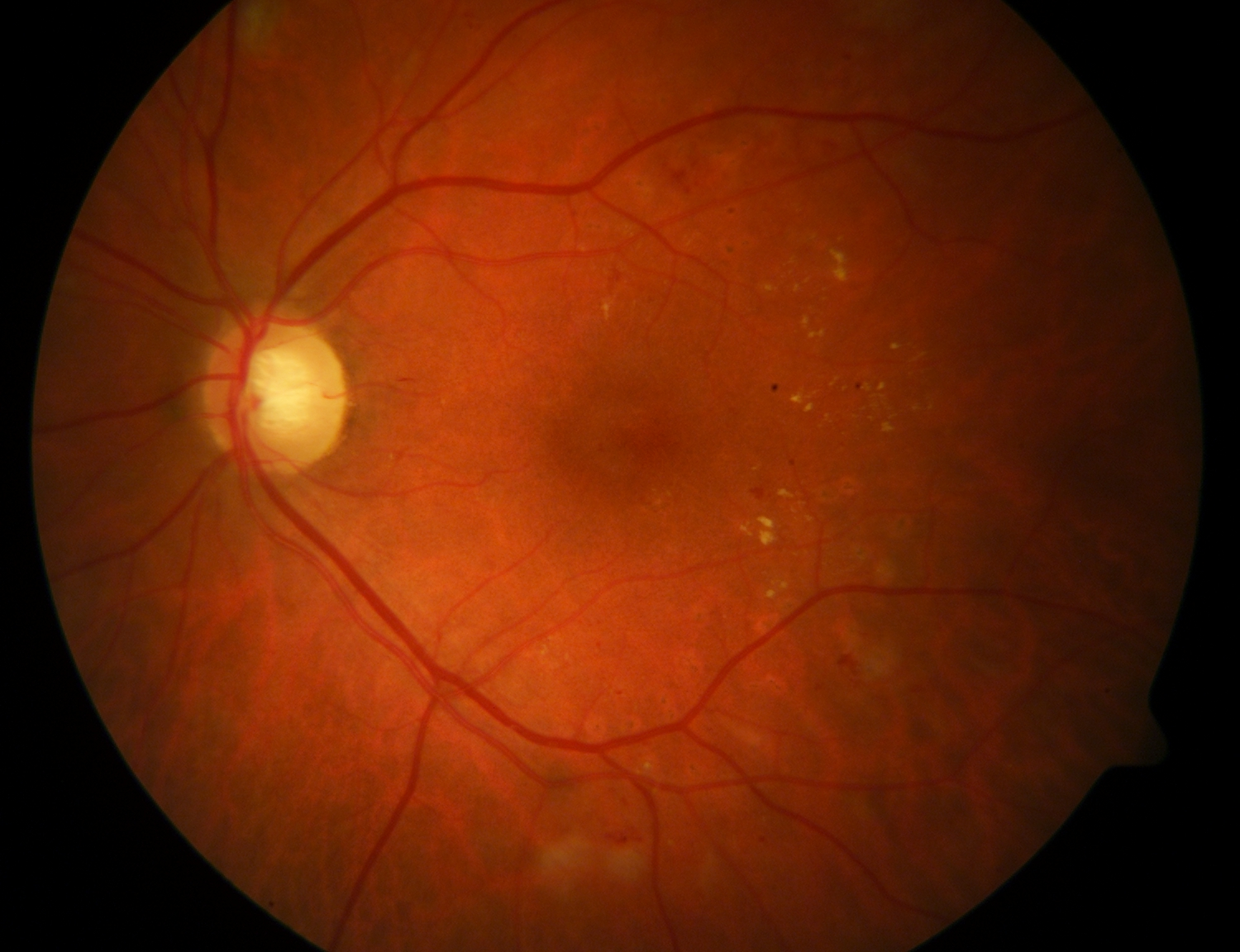}}
      \subfigure{\includegraphics[scale=0.075]{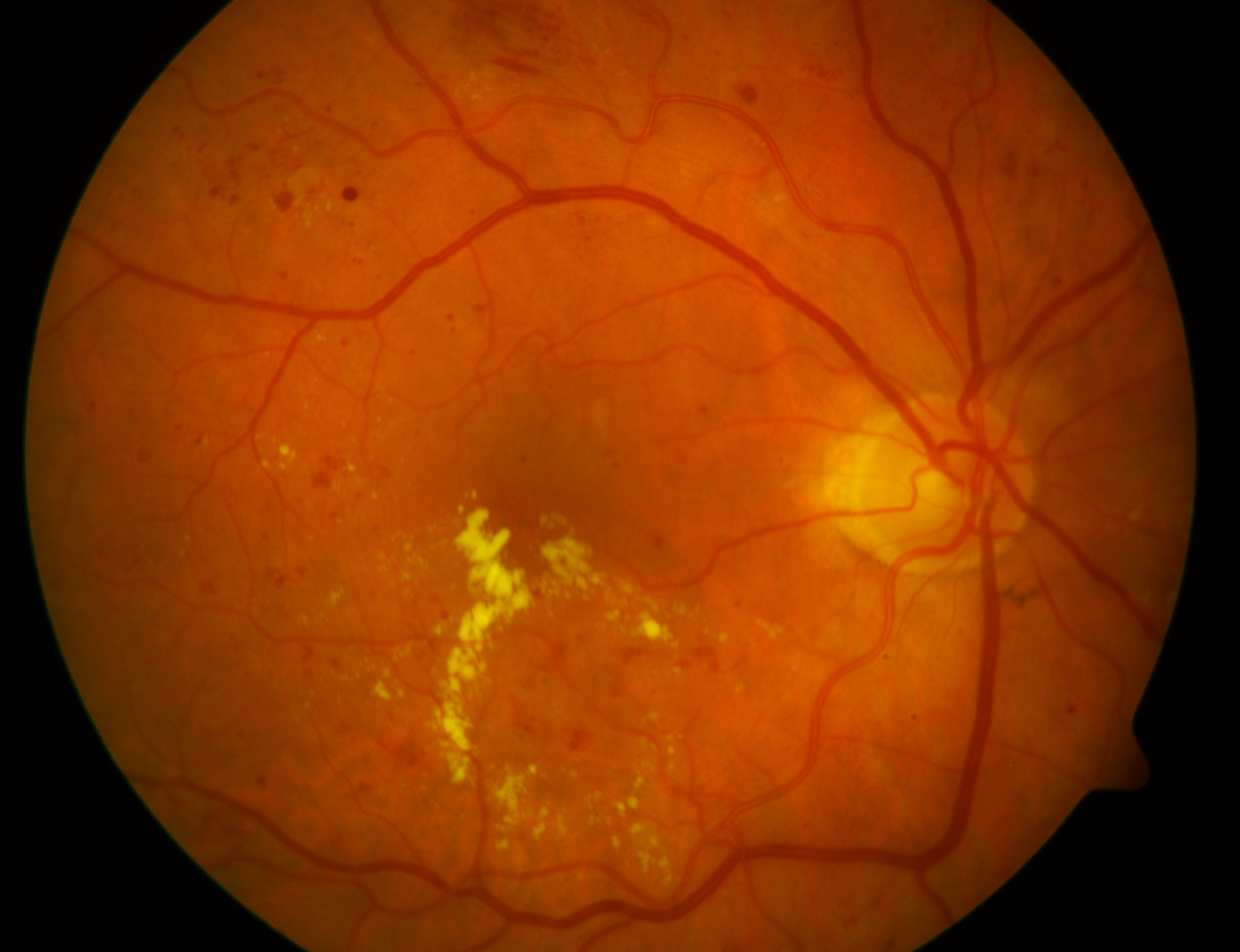}}
      \caption{Representative images(left(training),right(testing) from DIARETDB1. Segmentation algorithm used is based on multi-space clustering approach(\cite{inbook1}).}
                                \label{fig2}
\end{figure}

DIARETDB1\cite{misc1} database is considered which consists of fundus images with exudates of various sizes. This is a publicly available database comprising of 89 fundus images in which 48 contains signs of hard exudates as marked by the experts. As exudates are early signs of Diabetic Retinopathy, good segmentation algorithms which can segment even the very smallest exudate is required. It is also required to evaluate the segmentation approach by a metric which emphasizes the tiniest exudates. Here, we have obtained the segmented images using the multi-space clustering approach given by \cite{inbook1}. 15 images with varying sizes of exudates were hand-chosen for training and 33 random images for testing. Original images are of size 1500 pixels x 1152 pixels, and were down-sized to 800 pixels x 615 pixels, to enable the running of the segmentation algorithm. The largest exudate present is 1.41\%,and the smallest is 0.0002\% of the image. Few sample images in the database with exudates are shown in Figure.~\ref{fig2}.
\FloatBarrier

CBIS-DDSM database is considered which is a standardized version of DDSM database. The mammogram images are in DICOM format. U-net based segmentation approach\cite{unet1} is used to obtain the segmentation results for calcifications. 108 images were considered for the evaluation of the proposed metric. The images were cropped to remove the majority black pixels in the background for U-net segmentation. First column in Figure~\ref{fig:calcif} shows sample mammogram images with calcifications. 

\begin{figure}
      \centering
      \subfigure{\includegraphics[width=3 cm]{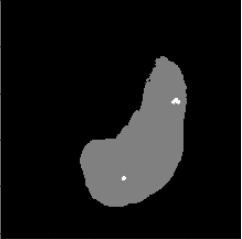}}
      \subfigure{\includegraphics[width=3 cm]{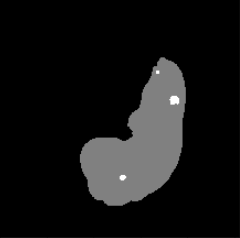}}
      \caption{Representative images of Liver\_tumor data set: left(ground truth),right(prediction).}
      \label{liv_dataset}
\end{figure}

131 abdomen images with various number of slices are taken from Medical Segmentation Decathlon challenge train data set (source: IRCAD Hôpitaux Universitaires)\cite{DBLP:journals/corr/abs-1902-09063} of Portal venous phase CT modality with liver and tumor as targets for segmentation. The data set is rich in terms of the challenges and variability of the images contained in it. The ground truth images in this data set is multi-labeled, where 0,1, and 2 are the respective labels for background, liver and tumor. The segmented images are obtained using the nnU-Net framework proposed by \cite{DBLP:journals/corr/abs-1904-08128}. Few sample slices with tumors are shown in Figure~\ref{liv_dataset}. Tumor appears in various sizes as seen in the figure.

107 abdomen images with various number of slices are taken from MSD challenge train data set (source: Memorial Sloan Kettering Cancer Center)\cite{DBLP:journals/corr/abs-1902-09063} of Portal venous phase CT modality with segmentation targets as pancreas and tumor. The data set contains cases with good variability. The segmented images are obtained using the nnU-Net framework proposed by \cite{DBLP:journals/corr/abs-1904-08128}. Here, only one tumor is present in a pancreas segment. In some slices, the pancreas tissue and tumor are of similar size, whereas in few slices, tumor is very small compared to the tissue leading to class imbalance.

%% file: expdis3.tex
The effectiveness of our metric in the evaluation of segmentation performance is demonstrated in this section. Results obtained on applying the proposed metric on each of the datasets has been has been reported. Comparisons with the existing widely-used metrics has also been presented. Besides, Hypothesis testing in order to establish the statistical significance of the proposed metric is also carried out.

\subsection{Experiments with synthetic images}
\begin{figure}
    \centering
    \subfigure[Ground truth image]{\includegraphics[scale=0.15]{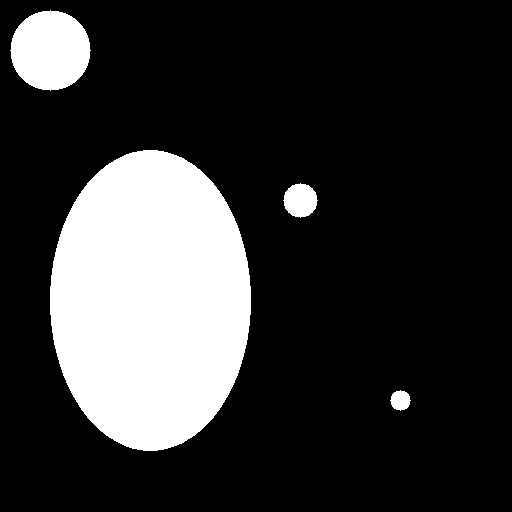}\label{sub:a}}
    \subfigure[Segmentation result from algorithm-1]{\includegraphics[scale=0.15]{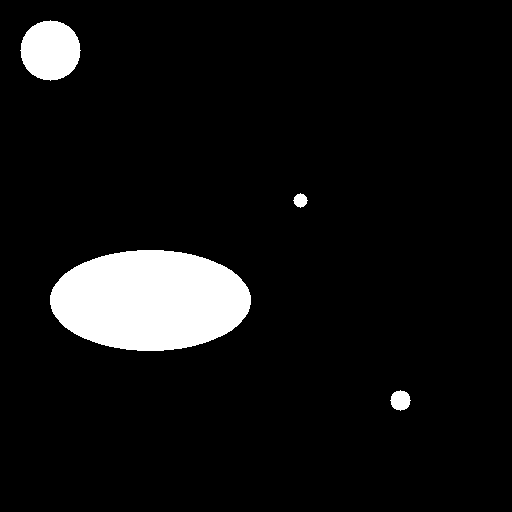}\label{sub:b}}\\
    \subfigure[Segmentation result from algorithm-2]{\includegraphics[scale=0.15]{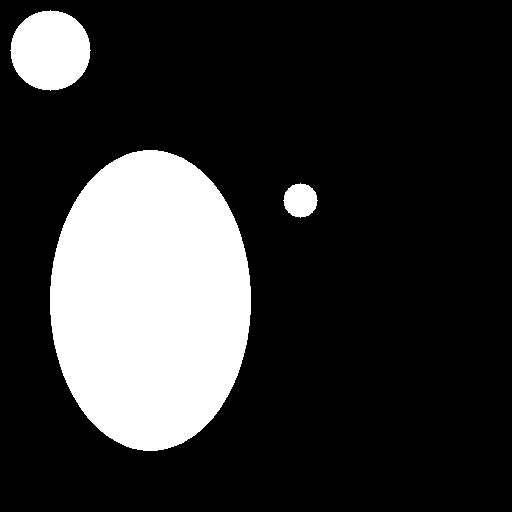}\label{sub:c}}
    \subfigure[Segmentation result from algorithm-3]{\includegraphics[scale=0.15]{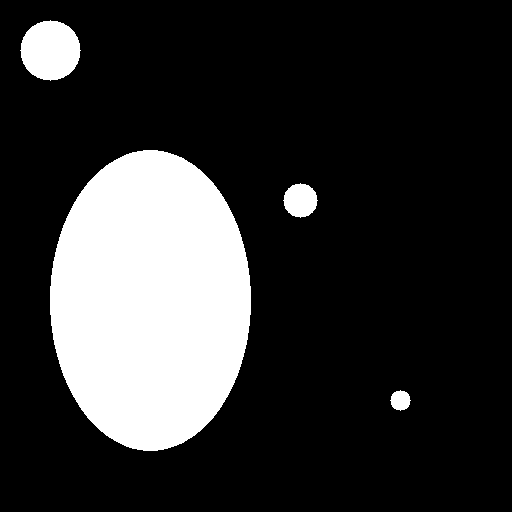}\label{sub:d}}
    \caption{A sample synthetic ground truth image(a) and predictions(b-d) with different segmentation algorithms to illustrate the proof of concept. Calculated metric values are given by: (b) Dice=0.5274, SSEGEP=0.5157, IOU=0.3582, Accuracy=0.8838 (c) Dice=0.9970, SSEGEP=0.75, IOU=0.9940, Accuracy=0.9987 (d) Dice=0.9790, SSEGEP=0.8903, IOU=0.9589, Accuracy=0.9916.}
    \label{synth_im_a}
\end{figure}

Few synthetic images with multiple segments of varying sizes which represent different segmentation scenarios are constructed as shown in Figure~\ref{synth_im_a}. The values of state-of-the-art metrics wrongly indicate that the result of second segmentation(~\subref{sub:c}) is better than that of third segmentation(\subref{sub:d})(It is because the number of detected exudate pixels with second algorithm is greater than  that  of  third  algorithm). But  a small exudate is missed in \subref{sub:c}, which means the corresponding segmentation algorithm is not efficient in detecting smaller exudates. The proposed metric can discriminate these cases as it weighs the \textit{TP} pixels based on the area of the segment it is derived from. \textit{SSEGEP} for \subref{sub:d} is higher than \subref{sub:c} which indicates that quality of prediction is better with algorithm 3 than algorithm 2.

\subsection{Experiments with real scan images}
\subsubsection{Experiments with retina images-Multiple segments of same type}
\begin{figure}
      \centering
      \includegraphics[scale=0.5]{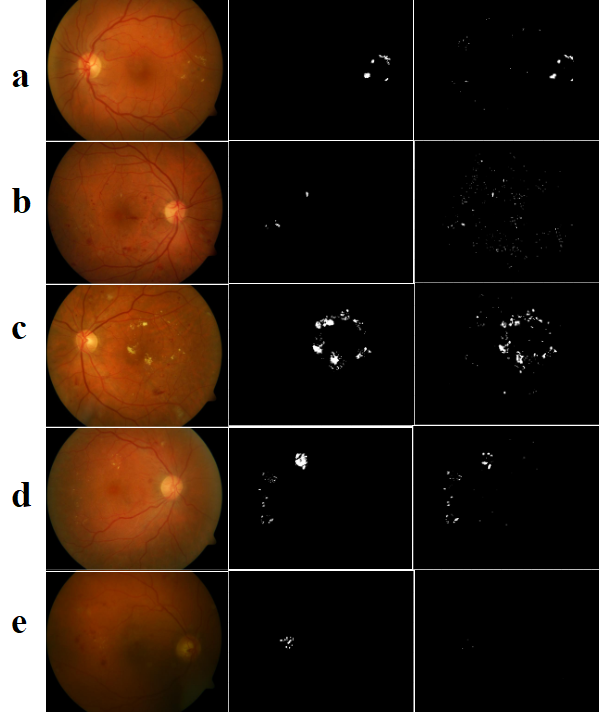}
      \caption{First column represents the original fundus images. Corresponding ground truths and predictions(using multi-space clustering) is shown in second and third column respectively. Refer Table~\ref{tab:fundus_scores} for the values of evaluation metrics(TPR,Accuracy,Dice,IOU,SSEGEP) and MOS.}
      \label{fig3}
\end{figure}

Figure~\ref{fig3} shows few fundus images with their ground truth and segmentation. The values of proposed metric and commonly used state-of-the-art metrics are compared with the average Mean Opinion Score. True positive rate is not sufficient to evaluate segmentation performance as it does not take into account the false positives(High \textit{TPR} in row 2 which is a poor segmentation). \textit{Dice} value can be used when there is no much variation in exudate size(rows 1,2,5). But \textit{Dice} fails to quantify segmentation performance when there are exudates of varying sizes(row 4). In such a case, \textit{SSEGEP} outperform other metrics as it is capable to discriminate different segments according to the count of the pixels in it.
\begin{table}[h!]
\tbl{A comparison of commonly used evaluation metrics with the proposed metric for fundus images in Figure~\ref{fig3} with exudates. Mean Opinion Score(MOS) is given in the last row.\label{tab:fundus_scores}}{%
\begin{tabular}{l  l   l   l   l   l   l}
\toprule
Panel & Sensitivity & Acc & Dice & IOU & SSEGEP & MOS \\
\colrule
a & 0.66 & 0.73 & 0.68 & 0.52 & \textbf{0.63} & \textbf{0.75}\\
b & 0.63 & 0.57 & 0.22 & 0.12 & \textbf{0.20} & \textbf{0.20}\\
c & 0.63 & 0.37 & 0.68 & 0.51 & \textbf{0.75} & \textbf{0.76}\\
d & 0.39 & 0.51 & 0.53 & 0.36 & \textbf{0.82} & \textbf{0.81}\\
e & 0.02 & 0.80 & 0.04 & 0.02 & \textbf{0} & \textbf{0.03}\\
\botrule
\end{tabular}}
\end{table}

The average of the difference between mean opinion score and value of the evaluation metric(SSEGEP,Dice and Sensitivity) are tabulated in Table~\ref{tab:tab1}. The deviation value is least with the proposed metric which indicates its higher correlation with human interpretability. 

\begin{table}[ht!]
\tbl{Comparison of proposed metric with the state of the art metrics on the
DIARETDB1 database is shown in second column.Average deviation of few evaluation
metrics from the mean opinion score evaluated by 15 subjects is given in column 3.
 Comparison is done on a scale of 0 to 1.\label{tab:tab1}}{%
\begin{tabular}{l l l}
\toprule
Evaluation  & Average Value  & Average deviation \\
Metric & obtained(Pixel level) & wrt to MOS \\
\colrule
Sensitivity & 0.4003 & 0.12\\
Dice value & 0.3539 & 0.11\\
SSEGEP(proposed) & \textbf{0.3324} & \textbf{0.08}\\
\botrule
\end{tabular}}
\end{table}

\subsubsection{Experiments with mammogram images-Multiple segments of same type}
\begin{figure}
      \centering
      \includegraphics[scale=0.2]{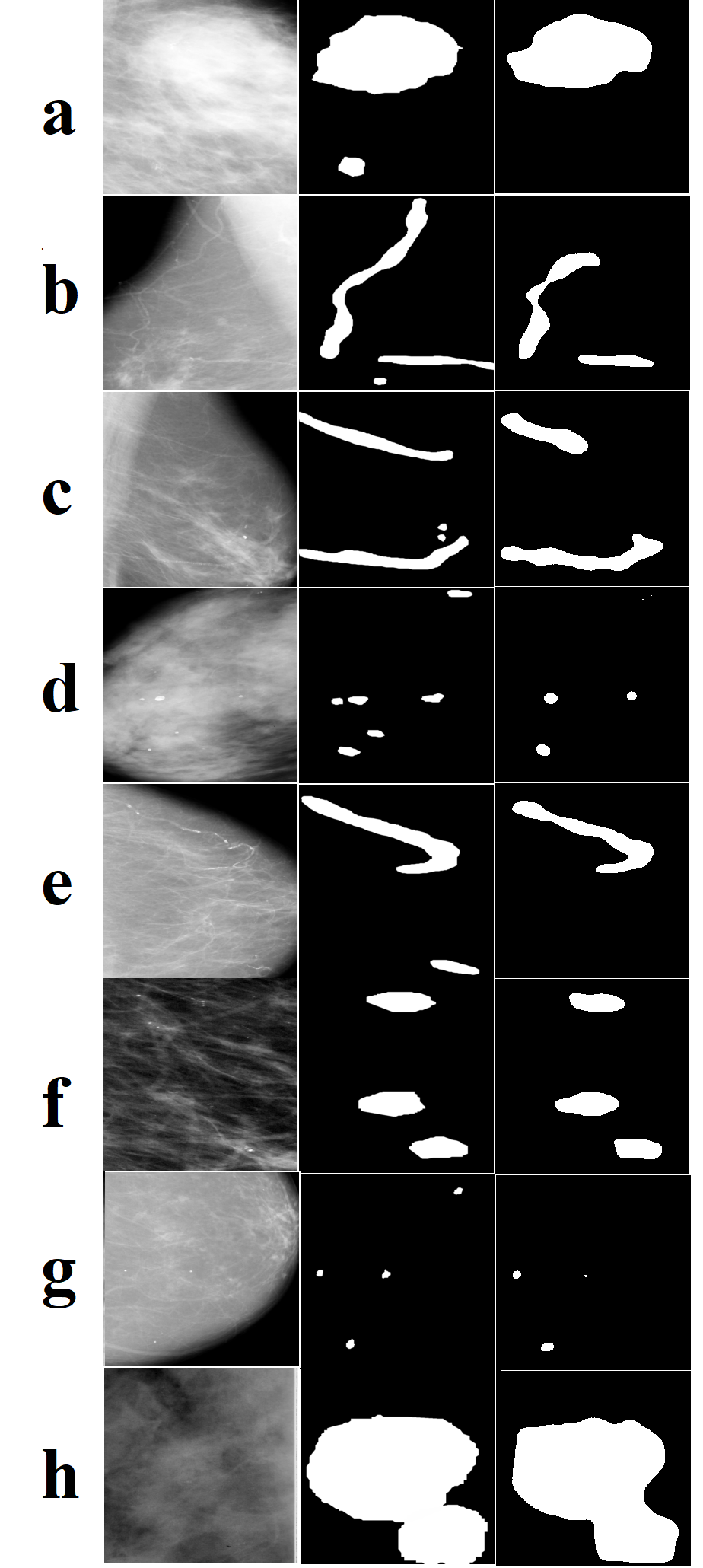}
      \caption{First column represents the original mammogram images. Corresponding ground truths and predictions(using U-net) is shown in second and third column respectively. Refer Table~\ref{Tab:calcif} for the values of evaluation metrics(TPR,Accuracy,Dice,IOU,SSEGEP,MCC).}
      \label{fig:calcif}
\end{figure}

Figure~\ref{fig:calcif} shows sample fundus images with their ground truth and segmentation masks. Table~\ref{Tab:calcif} shows the values of proposed metric and other commonly used metrics. Here, comparison is also made with another evaluation metric, \textit{MCC}(Matthews correlation coefficient). As seen in the fundus images, sensitivity and accuracy are not sufficient to quantify the segmentation performance. Panels 'a' to 'e' consists of calcifications of varying sizes. Smaller calcifications indicate the higher severity of disease.  Even though smaller segment is lost in mammogram segmentation for those panels, state-of-the-art metrics fail to reflect it. For eg., the \textit{Dice} value is 0.91 for mammogram in panel 'a' even though the segmentation algorithm failed to detect the smaller segment. But, \textit{SSEGEP} is 0.43 which indicates the poor segmentation in terms of medical diagnosis. Panels 'f' and 'g' consists of calcifications of similar sizes, hence the values of proposed metric, \textit{SSEGEP} is comparable to state-of-the-art metrics. Panel 'h' comprises of a single large segment, and hence all metric values are of similar range.
\begin{table}[ht!]
\tbl{A comparison of commonly used evaluation metrics with the proposed metric for mammogram images in Figure~\ref{fig:calcif} with calcification.\label{Tab:calcif}}{%
\begin{tabular}{l  l   l   l   l   l   l}
\toprule
Panel & Sensitivity & Acc & Dice & IOU & MCC & SSEGEP \\
\colrule
a & 0.84 & 0.96 & 0.91 & 0.84 & 0.88 & \textbf{0.43}\\
b & 0.59 & 0.95 & 0.68 & 0.52 & 0.66 & \textbf{0.39}\\
c & 0.78 & 0.95 & 0.76 & 0.61 & 0.73 & \textbf{0.59}\\
d & 0.35 & 0.99 & 0.49 & 0.32 & 0.53 & \textbf{0.29}\\
e & 0.70 & 0.97 & 0.81 & 0.68 & 0.81 & \textbf{0.40}\\
f & 0.83 & 0.98 & 0.89 & 0.81 & 0.89 & \textbf{0.81}\\
g & 0.48 & 0.99 & 0.54 & 0.37 & 0.54 & \textbf{0.42}\\
h & 0.89 & 0.94 & 0.94 & 0.88 & 0.89 & \textbf{0.88}\\
\botrule
\end{tabular}}
\end{table}

\begin{figure}
      \centering
      \includegraphics[scale=0.6]{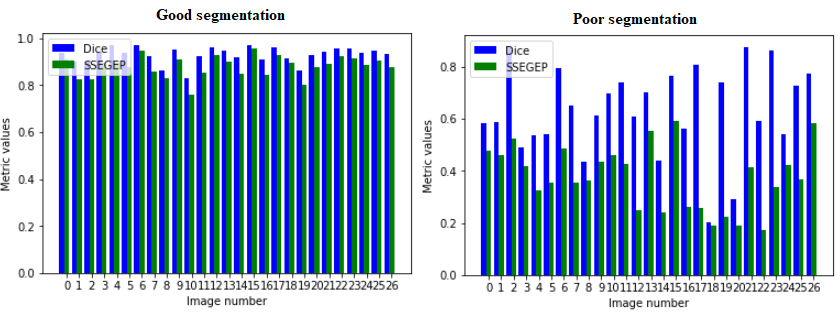}
      \caption{Bar plots for \textit{SSEGEP} and \textit{Dice} values for 'good' and 'bad' segmentation categories(as mentioned in 'Hypothesis Testing' in section 4). Metric values are comparable for images in good segmentation group. But the Dice values are greater than SSEGEP for images under poor segmentation category.}
      \label{fig:scatterplot}
        \vspace{-2 em}
\end{figure}

The bar plots of SSEGEP values and Dice values for few samples of images are shown in Figure~\ref{fig:scatterplot}. Dice and SSEGEP values for images of good segmentation category are almost in the same range. It implies that both the metrics are able to reflect good segmentations. But the Dice values for many images under poor segmentation category are higher than that of SSEGEP values. Even though the segmentation result is poor, Dice values are high. But the proposed metric is able to reflect the poor quality of segmentation, since those values are lower. 

The results for statistical significance test are tabulated in Table~\ref{tab:tab2}. The p-value for \textit{SSEGEP} is lesser than that of \textit{Dice} value, which indicates the better discriminating capability for the proposed metric.

\subsubsection{Experiments with Liver images}
\begin{figure}
      \centering
      \includegraphics[scale=0.4]{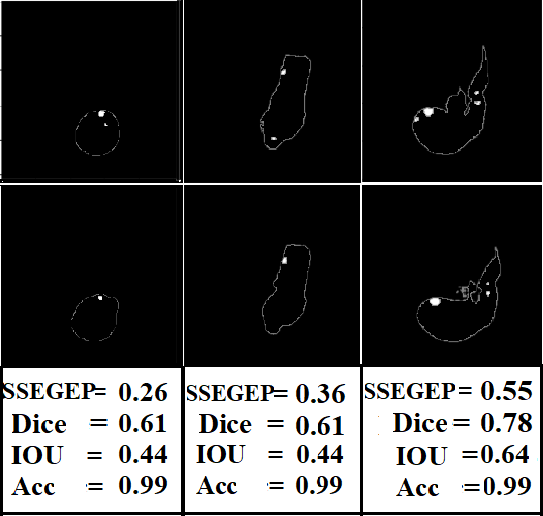}
      \caption{Performance evaluation of tumor segmentation(using nnU-Net) of liver(all tumors are smaller in size) is illustrated in the figure. First row represents the ground truth images. Corresponding segmentation is shown in the second row. The values of evaluation metrics(only for tumor): SSEGEP, Dice, IOU and accuracy is given in the last row.}
      \label{figsm}
        \vspace{-2.0em}
\end{figure}

\begin{figure}
      \centering
      \includegraphics[scale=0.4]{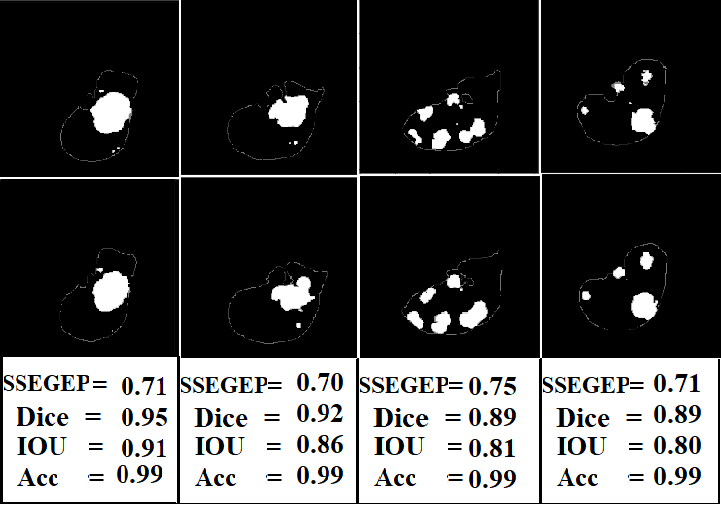}
      \caption{Performance evaluation of tumor segmentation(using nnU-Net) of liver(consists of larger and smaller tumors) is illustrated in the figure. First row represents the ground truth images. Corresponding segmentation is shown in the second row. The values of evaluation metrics(only for tumor): SSEGEP, Dice, IOU and accuracy is given in the last row.}
      \label{figmix}
        \vspace{-1.5em}
\end{figure}

\begin{figure}
      \centering
      \includegraphics[scale=0.4]{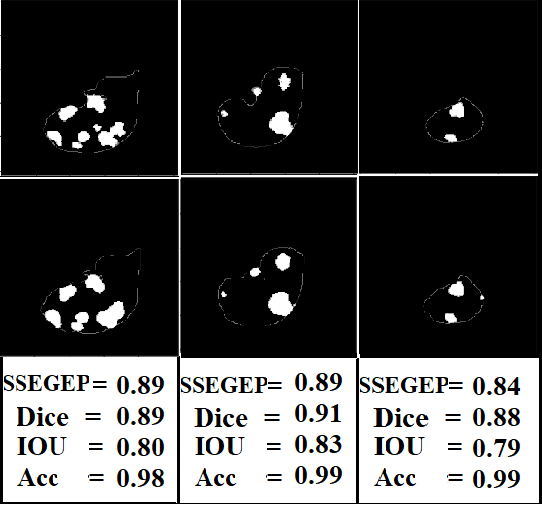}
      \caption{Performance evaluation of tumor segmentation(using nnU-Net) of liver(similar and moderately sized tumors) is illustrated in the figure. First row represents the ground truth images. Corresponding segmentation is shown in the second row. The values of evaluation metrics(only for tumor): SSEGEP, Dice, IOU and accuracy is given in the last row.}
      \label{figmod}
        \vspace{-1.5em}
\end{figure}

\paragraph{Multiple segments(tumors) of same type}
Figure~\ref{figsm},\ref{figmix} and \ref{figmod} shows the evaluation metrics obtained for few sample cases of (1) images with all small tumors, (2) images with small and large tumors and (3) images with similar sized(medium sized) tumors. Here, performance evaluation is conducted only for tumors. \textit{Accuracy} for all the cases is high(0.98,0.99) as the number of \textit{TN} pixels is high compared to other pixels. As seen in Figure~\ref{figsm}, all the tumors appear small. \textit{SSEGEP} and \textit{IOU} values are varying in a similar way. \textit{Dice} values are slightly higher than \textit{IOU} as the former gives higher weightage to \textit{TP} pixels. Since the tumors are smaller in size, even though few pixels are missed in segmentation, it should be highly penalized which is reflected by the proposed metric.

Figure~\ref{figmix} consists of tumors of varying sizes. There are tiny tumors and large tumors present in a single tissue. So, the loss of a small tumor should be highly penalized. But still the values of other metrics are closer to 0.90. However, \textit{SSEGEP} value is in the range of 0.70-0.75, thereby reflecting the level of quality of segmentation.

Figure~\ref{figmod} comprises of scan images with similar and moderately sized tumors. The \textit{SSEGEP} values are comparable with other commonly used metrics because of similar sized segments. There is a \textit{FP} present in the third case which resulted in lesser \textit{SSEGEP} value compared to other two cases. Since weightage of \textit{FP} is decided by the inverse of \textit{TP} pixels and number of \textit{TP} pixels is less in this case, \textit{SSEGEP} value is not too low. 

\begin{figure}
      \centering
      \includegraphics[scale=0.4]{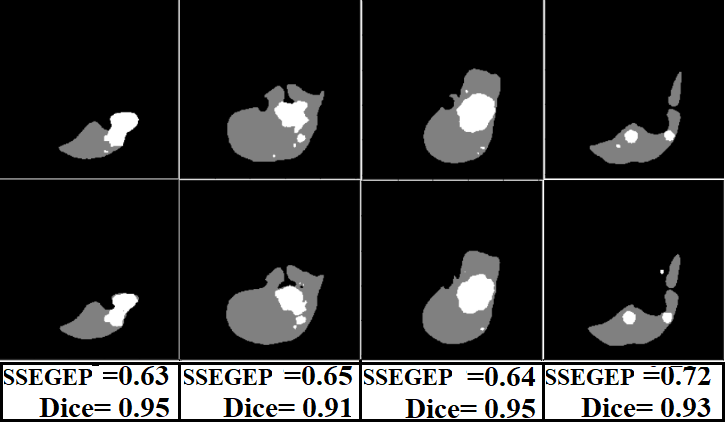}
      \caption{Combined segmentation(Using nnU-net) of liver and tumor with the evaluation metrics is given.First row represents the ground truth images. Corresponding segmentation is shown in the second row. The values of evaluation metrics: SSEGEP and Dice are given in the last row.}
      \label{fig6}
        \vspace{-2.0em}
\end{figure}

Hence, we can conclude that if there are tumors of varying sizes in ground truth and if the segmentation algorithm fails to extract the smaller tumors, it should be severely penalized (as seen in Figure~\ref{figmix}). As it can be clearly seen in Figure~\ref{figmix}, even though other metric values are at a higher side indicating best segmentation, proposed metric is lesser than that since smaller tumors are lost in segmentation. Commonly used metrics fails to discriminate the varying sized distinct tumors. But the proposed metric assigns different weightage to tumors based on their size, emphasizing smaller ones. 
\paragraph{Multi-label case with many segments of different types(liver and tumor)with class imbalance}
Figure~\ref{fig6} shows sample multi-label segmentation and the corresponding evaluation metrics. Here, \textit{SSEGEP} is compared with the generalized multi class \textit{Dice} value which is the commonly used performance metric for unbalanced multi-class segmentation. Even though generalized multi class \textit{Dice} can discriminate the two different labels, it fails to differentiate the multi-sized tumors present.
As seen in the figure~\ref{fig6}, liver occupies more space compared to the tumors. Even though small tumors are not identified in each case, \textit{Dice} value is same as it gives the same weightage to every tumor irrespective of their size. But \textit{SSEGEP} value is not high unlike \textit{Dice} value since the proposed metric emphasizes the smallest tumor more by assigning a higher weightage to it, thereby penalizing the loss of smaller segments.
The results for hypothesis testing are tabulated in Table~\ref{tab:tab2}. The test says that the inputs come from different distributions for both the metrics with p-value for \textit{SSEGEP} lesser than that of \textit{Dice} value. This implies that both the metrics(\textit{SSEGEP} and \textit{Dice}) are able to distinguish between good and poor segmentation, with a better discriminating capability for \textit{SSEGEP}.

\subsubsection{Experiments with pancreas images-Multi-label case with one segment each of different type(pancreas and tumor)}
\begin{figure}
      \centering
      \includegraphics[scale=0.4]{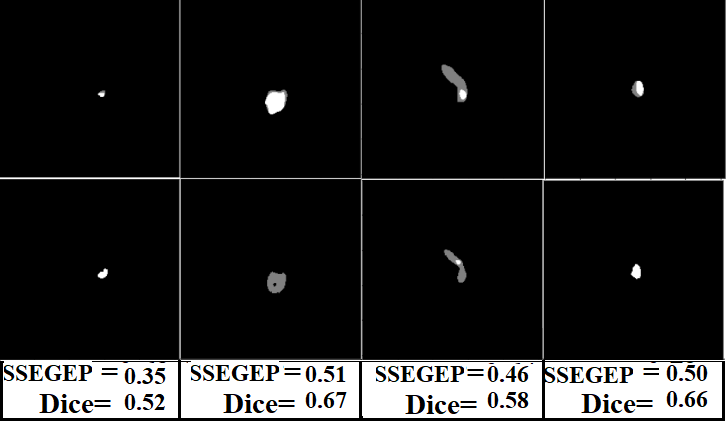}
      \caption{Combined segmentation(Using nnU-net) of pancreas and tumor with the evaluation metrics is given.First row represents the ground truth images. Corresponding segmentation is shown in the second row. The values of evaluation metrics: SSEGEP and Dice are given in the last row.}
      \label{fig7}
        \vspace{-2.5em}
\end{figure}

Here, only one tumor is present in a pancreas segment. So, the evaluation is done for combined pancreas and tumor image as explained for liver and tumor case. Sample images with generalized multi class \textit{Dice} and \textit{SSEGEP} values are shown in Figure~\ref{fig7}. Tumor is not detected for ground truth in the first and last column in Figure~\ref{fig7}. For all cases, \textit{Dice} and \textit{SSEGEP} values vary in similar way, since each class(pancreas and tumor) contains only one segment(one tissue and one tumor) unlike combined liver and tumor case(in liver and tumor segmentation, there were multiple tumors in the liver tissue). Hence, for multi label segmentation evaluation where there is only one segment present in a class(or label), either generalized \textit{Dice} or \textit{SSEGEP} can be used.  

The results for hypothesis testing are tabulated in Table~\ref{tab:tab2}. The p-value is same for both metrics, indicating the applicability of both metrics in that case. This is because there is only one segment corresponding to each label in an image slice and both \textit{SSEGEP} and generalized \textit{Dice} acts in similar way for that scenario, whereas in liver tumor segmentation there are tumors of various sizes and the size variation within a label is not reflected by \textit{Dice} value.
\subsection{Analysis of impact of parameter changes on SSEGEP}
The various parameters used in the study are:
\begin{enumerate}
\item Number of connected components in the Ground Truth: If the number of connected components in the Ground Truth, as an extreme case, happens to be only 1, then the proposed metric would not result in any significantly different way compared to \textit{Dice} score, since it simply involves the calculation of total overlap area like \textit{Dice} score.
\item The Relative sizes of each of the connected components in the Ground Truth: If the sizes of each of the connected components in the Ground Truth are all comparable, then the advantage of assigning them different weights will not lead to much distinction across them. However, if the sizes of the connected components in the Ground Truth are very diverse, then the significance of assigning differential weights to each of them becomes very perceptible. 
\item Updated Weightage for \textit{TP}: Weightage assigned to \textit{TP} depends on the size of the corresponding connected component in groundtruth. If the connected component is of smaller size, higher weightage is assigned to that \textit{TP} pixel, and vice-versa.
\item Updated Weightage for \textit{FP}: Weightage assigned to \textit{FP} depends on the size of total \textit{TP} pixels. When the number of \textit{FP} pixels is less, the segmentation performance is better and when \textit{FP} is more, segmentation performance is poor. To weight the \textit{FP} pixels, inverse of \textit{TP} pixels is considered. Hence if the \textit{FP} count is higher than \textit{TP} count, a higher weightage is assigned to \textit{FP} thereby leading to smaller value of \textit{SSEGEP}.
\end{enumerate}

\begin{table}[h!]
\tbl{Hypothesis Testing with $\alpha=0.00001$.\label{tab:tab2}}{%
\begin{tabular}{l l l}
\toprule
Segmentation scenarios & p value(Dice) & p value(SSEGEP)  \\
\colrule
Calcification & 1.8e-19 & 3.5e-35\\
Tumor(liver) & 7.6e-12 & 1.2e-18\\
Combined liver-tumor & 2.9e-17 & 3.6e-19\\
Combined pancreas-tumor & 3.3e-19 & 3.3e-19\\
\botrule
\end{tabular}}
\end{table}
\paragraph{Limitations of proposed metric}
The above experiments say that proposed metric gives a better performance than other metrics for the evaluation of segmentation performance of medical images with segments of varying sizes. For images with segments of similar size, \textit{SSEGEP} performs in similar way as other overlap based metrics(\textit{Dice} score,\textit{IOU}). If the segmentation is done in random or by chance, the metric value should ideally be zero. But this is not the case with \textit{SSEGEP}.

%% file: JMLFS_template.bbl
\begin{thebibliography}{10}

\bibitem{Paper6}
A.~{Mason}, J.~{Rioux}, S.~E. {Clarke}, A.~{Costa}, M.~{Schmidt}, V.~{Keough},
  T.~{Huynh}, and S.~{Beyea}.
\newblock Comparison of objective image quality metrics to expert
  radiologists’ scoring of diagnostic quality of mr images.
\newblock {\em IEEE Transactions on Medical Imaging}, pages 1--1, 2019.

\bibitem{p10}
Kelly Zou, Simon Warfield, Aditya Bharatha, Clare Tempany, Michael Kaus, Steven
  Haker, William Wells, Ferenc Jolesz, and Ron Kikinis.
\newblock Statistical validation of image segmentation quality based on a
  spatial overlap index.
\newblock {\em Academic radiology}, 11:178--89, 2004.

\bibitem{p11}
KH~Zou, WM~Wells, R~Kikinis, and SK~Warfield.
\newblock Three validation metrics for automated probabilistic image
  segmentation of brain tumours.
\newblock {\em Stat Med}, 23:1259--82, 2004.

\bibitem{p12}
Andrew Worth, Makris Nikos, Verne Caviness, and David Kennedy.
\newblock Neuroanatomical segmentation in mri: Technological objectives.
\newblock {\em International Journal of Pattern Recognition and Artificial
  Intelligence}, 11:1161–87, 11 1997.

\bibitem{p13}
SK~Warfield, CF~Westin, CRG Guttmann, MS~Albert, FA~Jolesz, and R~Kikinis.
\newblock Fractional segmentation of white matter.
\newblock In {\em International Conference on Energy Efficient Technologies for
  Sustainability (ICEETS)}, pages 109--117, 1999.

\bibitem{brainct}
A.~Nanthagopal and R~Sukanesh.
\newblock Combined texture feature analysis of segmentation and classification
  of benign and malignant tumor ct slices.
\newblock {\em Journal of medical engineering \& technology}, 37, 10 2012.

\bibitem{mri_breastim}
Mohammed El~Adoui, Sidi Mahmoudi, Amine Larhmam, and Mohammed Benjelloun.
\newblock Mri breast tumor segmentation using different encoder and decoder cnn
  architectures.
\newblock {\em Journal of Computers}, 8:52, 06 2019.

\bibitem{multidice}
Carole Sudre, Wenqi Li, Tom Vercauteren, Sebastien Ourselin, and Manuel~Jorge
  Cardoso.
\newblock Generalised dice overlap as a deep learning loss function for highly
  unbalanced segmentations.
\newblock In {\em Deep Learning in Medical Image Analysis and Multimodal
  Learning for Clinical Decision Support}, pages 240--248, 09 2017.

\bibitem{DR1}
R~Klein, B.E.K. Klein, S~E~Moss, M~D~Davis, and David Demets.
\newblock The wisconsin epidemiologic study of diabetic retinopathy. vii.
  diabetic nonproliferative retinal lesions.
\newblock {\em Ophthalmology}, 94:1389--400, 12 1987.

\bibitem{evidencefundus}
Prisca Loganadane, Bernard Delbosc, and Maher Saleh.
\newblock Short-term progression of diabetic hard exudates monitored with
  high-resolution camera.
\newblock {\em Ophthalmic Research}, pages 1--7, 11 2018.

\bibitem{exudates1}
Qing Liu, Beiji Zou, Jie Chen, Wei Ke, Kejuan Yue, Zailiang Chen, and Guoying
  Zhao.
\newblock A location-to-segmentation strategy for automatic exudate
  segmentation in colour retinal fundus images.
\newblock {\em Computerized Medical Imaging and Graphics}, 55, 09 2016.

\bibitem{exudates2}
Sreeparna Banerjee and Diptoneel Kayal.
\newblock Detection of hard exudates using mean shift and normalized cut
  method.
\newblock {\em Biocybernetics and Biomedical Engineering}, 36, 10 2016.

\bibitem{smallhepaticlesions}
Lawrence Schwartz, Eric Gandras, Sandra Colangelo, Matthew Ercolani, and David
  Panicek.
\newblock Prevalence and importance of small hepatic lesions found at ct in
  patients with cancer.
\newblock {\em Radiology}, 210:71--4, 02 1999.

\bibitem{disapp_cal}
Quan Nguyen, Nga Nguyen, Linden Dixon, Flavia Monetto, and Angelica Robinson.
\newblock Spontaneously disappearing calcifications in the breast: A rare
  instance where a decrease in size on mammogram is not good.
\newblock {\em Cureus}, 12, 06 2020.

\bibitem{calref1}
Benoit Mesurolle, Fawaz Halwani, Vincent Pelsser, Jean Gagnon, Ellen Kao, and
  Francine Tremblay.
\newblock Spontaneous resolving breast microcalcifications associated with
  breast carcinoma.
\newblock {\em The breast journal}, 11:478--9, 11 2005.

\bibitem{calref2}
H.~Seymour, J.~Cooke, and R.~Given-Wilson.
\newblock The significance of spontaneous resolution of breast calcification.
\newblock {\em The British journal of radiology}, 72 853:3--8, 1999.

\bibitem{calref3}
A~Evans, K~Clements, A~Maxwell, H~Bishop, A~Hanby, G~Lawrence, et~al.
\newblock Lesion size is a major determinant of the mammographic features of
  ductal carcinoma in situ: findings from the sloane project.
\newblock {\em Breast Cancer Research}, 65:181--4, 2010.

\bibitem{calref4}
Breastcancer.org.
\newblock {\em Understanding Breast Calcifications}, 2018 (accessed October 13,
  2020).

\bibitem{unet1}
Olaf Ronneberger, Philipp Fischer, and Thomas Brox.
\newblock U-net: Convolutional networks for biomedical image segmentation,
  2015.

\bibitem{keyvan}
Keyvan Kasiri, Mohammad Dehghani, Kamran Kazemi, Mohammad Helfroush, and
  S~Kafshgari.
\newblock Comparison evaluation of three brain mri segmentation methods in
  software tools.
\newblock In {\em 2010 17th Iranian Conference of Biomedical Engineering
  (ICBME)}, pages 1 -- 4, 12 2010.

\bibitem{paper1}
Ona Wu, Stefan Winzeck, Anne-Katrin Giese, Brandon Hancock, Mark Etherton, Mark
  Bouts, Kathleen Donahue, Markus Schirmer, Robert Irie, Steven Mocking, Elissa
  McIntosh, Raquel Bezerra, Konstantinos Kamnitsas, Petrea Frid, Johan
  Wasselius, John Cole, Huichun Xu, Lukas Holmegaard, Jordi Jimenez-Conde, and
  Natalia Rost.
\newblock Big data approaches to phenotyping acute ischemic stroke using
  automated lesion segmentation of multi-center magnetic resonance imaging
  data.
\newblock {\em Stroke}, 50, 06 2019.

\bibitem{paper2}
Gabriele Piantadosi, Mario Sansone, Roberta Fusco, and Carlo Sansone.
\newblock Multi-planar 3d breast segmentation in mri via deep convolutional
  neural networks.
\newblock {\em Artificial Intelligence in Medicine}, page 101781, 12 2019.

\bibitem{paper4}
Ge, Mu~Ting, Zhan Ning, Chen Tianming, Gao Zhi, Mu~Wanrong, and Shanxiang.
\newblock Multi-planar 3d breast segmentation in mri via deep convolutional
  neural networks.
\newblock {\em Computational intelligence and neuroscience}, 7 2019.

\bibitem{paper8}
N~Nur and Handayani Tjandrasa.
\newblock Exudate segmentation in retinal images of diabetic retinopathy using
  saliency method based on region.
\newblock {\em Journal of Physics: Conference Series}, 1108:012110, 11 2018.

\bibitem{paper9}
A~R Abblin and S~A Praylin Selva~Blessy.
\newblock Automatic segmentation and recognition of iris from an eye image.
\newblock In {\em International Conference on Energy Efficient Technologies for
  Sustainability (ICEETS)}, 2018.

\bibitem{paper10}
Ebenezer Priya and S.~Srinivasan.
\newblock Automated object and image level classification of tb images using
  support vector neural network classifier.
\newblock {\em Biocybernetics and Biomedical Engineering}, 36, 07 2016.

\bibitem{paper11}
Fang Lu, Fa~Wu, Peijun Hu, Zhiyi Peng, and Dexing Kong.
\newblock Automatic 3d liver location and segmentation via convolutional neural
  network and graph cut.
\newblock {\em International Journal of Computer Assisted Radiology and
  Surgery}, 12, 09 2016.

\bibitem{paper12}
Peter Rot, Matej Vitek, Klemen Grm, {\v{Z}}iga Emer{\v{s}}i{\v{c}}, Peter Peer,
  and Vitomir {\v{S}}truc.
\newblock {\em Deep Sclera Segmentation and Recognition}, pages 395--432.
\newblock Springer International Publishing, Cham, 2020.

\bibitem{Paper3}
Dennis Bontempi, Sergio Benini, Alberto Signoroni, Michele Svanera, and Lars
  Muckli.
\newblock Cerebrum: a fast and fully-volumetric convolutional encoder-decoder
  for weakly-supervised segmentation of brain structures from
  out-of-the-scanner mri.
\newblock {\em ArXiv}, abs/1909.05085, 2019.

\bibitem{paper7}
Christoph Baur, Benedikt Wiestler, Shadi Albarqouni, and Nassir Navab.
\newblock Fusing unsupervised and supervised deep learning for white matter
  lesion segmentation.
\newblock In {\em International Conference on Medical Imaging with Deep
  Learning -- Full Paper Track}, London, United Kingdom, 08--10 Jul 2019.

\bibitem{dice1}
Piotr Chudzik, Somshubra Majumdar, Francesco Caliva, Bashir Al-Diri, and Andrew
  Hunter.
\newblock {Exudate segmentation using fully convolutional neural networks and
  inception modules}.
\newblock In {\em Medical Imaging 2018: Image Processing}, volume 10574, pages
  785 -- 792. SPIE, 2018.

\bibitem{dice2}
Sylvain Gouttard, Martin Styner, Marcel Prastawa, Joseph Piven, and Guido
  Gerig.
\newblock Assessment of reliability of multi-site neuroimaging via traveling
  phantom study.
\newblock In Dimitris Metaxas, Leon Axel, Gabor Fichtinger, and G{\'a}bor
  Sz{\'e}kely, editors, {\em Medical Image Computing and Computer-Assisted
  Intervention -- MICCAI 2008}, pages 263--270, Berlin, Heidelberg, 2008.
  Springer Berlin Heidelberg.

\bibitem{paper5}
Abraham Nabila and Khan Naimul.
\newblock A novel focal tversky loss function with improved attention u-net for
  lesion segmentation.
\newblock In {\em 2019 IEEE International Symposium on Biomedical Imaging
  (ISBI)}, 2019.

\bibitem{familymetrics}
Varduhi Yeghiazaryan and Irina Voiculescu.
\newblock Family of boundary overlap metrics for the evaluation of medical
  image segmentation.
\newblock {\em Journal of Medical Imaging}, 5:1, 02 2018.

\bibitem{ref_hau}
Abdel~Aziz Taha and Allan Hanbury.
\newblock Metrics for evaluating 3d medical image segmentation: analysis,
  selection, and tool.
\newblock {\em BMC medical imaging}, 15:29, 2015.

\bibitem{mcc_ref}
Pierre Baldi, SÃ¸ren Brunak, Yves Chauvin, Claus Andersen, and Henrik
  Nielsen.
\newblock Assessing the accuracy of prediction algorithms for classification:
  An overview.
\newblock {\em Bioinformatics (Oxford, England)}, 16:412--24, 06 2000.

\bibitem{misc1}
V.~Kalesnykiene, J.~k.~Kamarainen, R.~Voutilainen, J.~Pietilä,
  H.~Kälviäinen, and H.~Uusitalo.
\newblock Diaretdb1 diabetic retinopathy database and evaluation protocol,
  2007.

\bibitem{inbook1}
Sanjeev Dubey and Utkarsh Mittal.
\newblock Exudate detection in fundus images: Multispace clustering approach.
\newblock In {\em Third International Conference, ICICCT 2018}, pages 109--117,
  01 2019.

\bibitem{DBLP:journals/corr/abs-1902-09063}
Amber~L. Simpson, Michela Antonelli, Spyridon Bakas, Michel Bilello, Keyvan
  Farahani, Bram van Ginneken, Annette Kopp{-}Schneider, Bennett~A. Landman,
  Geert J.~S. Litjens, Bjoern~H. Menze, Olaf Ronneberger, Ronald~M. Summers,
  Patrick Bilic, Patrick~Ferdinand Christ, Richard K.~G. Do, Marc Gollub,
  Jennifer Golia{-}Pernicka, Stephan Heckers, William~R. Jarnagin, Maureen
  McHugo, Sandy Napel, Eugene Vorontsov, Lena Maier{-}Hein, and M.~Jorge
  Cardoso.
\newblock A large annotated medical image dataset for the development and
  evaluation of segmentation algorithms.
\newblock {\em CoRR}, abs/1902.09063, 2019.

\bibitem{DBLP:journals/corr/abs-1904-08128}
Fabian Isensee, Jens Petersen, Simon A.~A. Kohl, Paul~F. J{\"{a}}ger, and
  Klaus~H. Maier{-}Hein.
\newblock nnu-net: Breaking the spell on successful medical image segmentation.
\newblock {\em CoRR}, abs/1904.08128, 2019.

\end{thebibliography}
